\documentclass[12pt]{article}
\usepackage[latin9]{inputenc}
\usepackage{amsmath}
\usepackage{amssymb}
\usepackage{graphicx}
\usepackage{placeins}
\usepackage{lineno}
\usepackage{titlesec}

\usepackage{color}

\titleformat{\subsection}[runin]{\bfseries}{}{0.0em}{}

\newcommand{\beq}{\begin{equation}}
\newcommand{\eeq}{\end{equation}} 
\newcommand{\beqa}{\begin{eqnarray}}
\newcommand{\eeqa}{\end{eqnarray}} 

\begin{document}

\title{Vortex lattice in the crossover of a Bose gas 
from weak coupling to unitarity}
\author{S.K. Adhikari$^{1}$ and L. Salasnich$^{2,3,*}$}

\date{\today}

\maketitle

\begin{center}
$^{1}$Instituto de F\'{\i}sica Te\'orica, UNESP - Universidade 
Estadual Paulista, 01.140-070 S\~ao Paulo, S\~ao Paulo, Brazil  
\\
$^{2}$Dipartimento di Fisica e Astronomia ``Galileo Galilei'', 
Universit\`a di Padova, Via Marzolo 8, 35131 Padova, Italy 
\\
$^{3}$Istituto Nazionale di Ottica (INO) del Consiglio Nazionale
delle Ricerche (CNR), Via Nello Carrara 1,
50019 Sesto Fiorentino, Italy
\\
$^{*}$Correspondence to luca.salasnich@unipd.it
\end{center}

\begin{abstract}
The formation of a regular lattice of quantized vortices 
in a fluid under rotation is a smoking-gun signature 
of its superfluid nature. { Here} we study the vortex 
lattice in a dilute { superfluid}  gas of bosonic atoms 
at zero temperature along the crossover from { the} weak-coupling regime, 
{ where the inter-atomic scattering length is very small 
compared to the average distance between atoms}, 
to { the} unitarity { regime}, where the 
inter-atomic scattering length diverges. 
This study is based on high-performance numerical simulations of the 
time-dependent nonlinear Schr\"odinger equation for the superfluid 
order parameter in three spatial dimensions, 
using  a realistic analytical expression for the bulk 
{ equation of state of the system} 
along the crossover from weak-coupling to unitarity.
{ This equation of state} has the correct weak-coupling 
and unitarity limits and faithfully reproduces the results of 
an accurate { multi-orbital} microscopic calculation. 
Our numerical predictions { of} the number of vortices and 
 root-mean-square sizes are 
important benchmarks for  future experiments. 
\end{abstract}

{At ultralow temperatures a dilute gas of bosonic atoms undergoes
a phase transition to a superfluid state of matter known as a Bose-Einstein
condensate. Soon after the observation of Bose-Einstein condensates 
in alkali-metal atoms \cite{becex}, experiments revealed \cite{vl}
the formation of vortices  in the form of triangular lattice 
in a rapidly rotating Bose-Einstein condensate 
demonstrating its superfluid nature.}
{ For such} dilute quantum gases 
the effective range of the  { inter-atomic} interaction is much smaller 
than the average distance between two atoms. Under this condition 
the interaction { can be} characterized by  { a single} 
parameter, the so-called s-wave { atomic}   scattering length 
$a$  \cite{leggett,rmp} { taken to be positive (repulsive interaction) 
throughout this study.  Most experiments on Bose-Einstein condensates 
were performed in the weak-coupling limit characterized by  small values 
of the gas parameter $x\equiv n^{1/3}a$ ($x\ll 1$) \cite{rmp}, where $n$ 
is the density of the gas.  }{ The theoretical description of such 
a system is based on a mean-field nonlinear 
equation, known as the  Gross-Pitaevskii equation \cite{gross}, where 
the nonlinearity is determined by \cite{rmp} the 
bulk chemical potential of a uniform gas written as a function of
the density $n$ and the scattering length $a$.}

{ Quite remarkably, the scattering length $a$ { can now be} 
routinely { manipulated by  varying}  an external magnetic field 
near a Feshbach resonance,} { thus changing effectively the 
inter-atomic  interaction \cite{feshbach}.} As the scattering length 
$a$ becomes much larger than all length scales { of the system 
in the strong-coupling regime characterized by large values of the gas 
parameter ($x\gg 1$)} the system exhibits universal 
behavior \cite{gior,univ,castin} determined { only} by 
the density $n$ of the gas. The unitary limit, where $x\to +\infty$, 
{ can be achieved by increasing the scattering length to divergent 
values near a Feshbach resonance and} is characterized by simple 
universal laws arising from scale invariance. This limit is of great 
interest { in} fields as diverse as ultra-cold atoms \cite{univ,castin}, 
quark-gluon plasma \cite{quark}, neutron \cite{star} and 
Bose \cite{bstar} stars, superconductivity \cite{string}, and 
string theory \cite{asoke}. The unitary Fermi 
gas made of fermionic alkali-metal atoms has been largely 
investigated  \cite{zwerger} both experimentally and theoretically. 
In the case of bosonic atoms there are technical difficulties 
associated with a large three-body loss by molecule formation 
as the unitarity limit is approached by increasing the scattering length 
$a$ \cite{salomon,hadzibabic}. Only recently these difficulties have 
been overcome to observe a unitary Bose gas \cite{jin}. 

In the commonly studied weak-coupling limit ($x \ll 1$), 
the zero-temperature bulk chemical potential $\mu (n,a)$ of a 
uniform three-dimensional (3D) Bose gas  is given by 
\beq
\mu(n,a)=4\pi n a + 2\pi \alpha n^{3/2} a^{5/2} + ... , 
 \quad \alpha = {64\over 3\sqrt{\pi}} \; , 
\label{lhy}
\eeq
where the first term is the familiar mean-field result, 
while the second term { is} the { beyond-mean-field}
Lee-Huang-Yang correction \cite{lee} which becomes important 
as $a$ increases. { Equation (\ref{lhy})} {  is the zero-temperature 
equation of state for a weakly-interacting { Bose} gas, 
containing the mean-field contribution and also the first 
beyond-mean-field correction} (for a recent review 
see \cite{sala-review}). Here we use 
$\hbar=m=1$ with $\hbar$ the reduced Planck constant and $m$ the 
mass of a boson. By dimensional arguments, 
the bulk chemical potential at unitarity is instead 
proportional to $n^{2/3}$, namely \cite{salomon,hadzibabic,jin}
\beq
\label{unit}
\lim _{a\to \infty} \mu (n,a)= \eta \ n^{2/3} \; , 
\eeq
where $\eta$ is a universal parameter. 
Although there is no experimental estimate of the parameter $\eta$ 
for bosons, there are some recent many-body calculations.
Ding and Greene (DG) \cite{DG} performed 
a microscopic { multi-orbital} calculation 
of $\mu(n,a)$ along  the crossover and obtained $\eta=4.7$. 
Also other theoretical predictions for $\eta$ \cite{SZ,LL,stoof,ZM,RS} 
lie in the range from $3$ to $9$, except $\eta =22.22$  of Ref. \cite{cowell}. 
In this  study we suggest  an analytic expression  for the 
zero-temperature bulk chemical potential $\mu(n,a)$, which reduces to 
Eqs. (\ref{lhy}) and (\ref{unit}) in the appropriate limits and which 
faithfully reproduces the results of the recent { multi-orbital} 
calculation by Ding and Greene \cite{DG}.  

{ From Eqs. (\ref{lhy}) and (\ref{unit}), we assume that the}
zero-temperature bulk chemical potential $\mu(n,a)$ along crossover 
from weak coupling to unitarity can be written as 
\beq
\label{eq2}
\mu (n,a) =  n^{2/3} f(x), \quad x \equiv n^{1/3}a \; ,  
\eeq
where $f(x)$ is a dimensionless  universal function. 
{ In this paper we propose a} 
{\it parameter-free minimal analytical} expression for 
the crossover function $f(x)$, 
incorporating the weak-coupling regime, Eq. (\ref{lhy}), and the  
the unitarity limit, Eq. (\ref{unit}), given by 
\beq
\label{co}
f(x)= 4\pi \frac{x+\alpha x^{5/2}}{1+\frac{\alpha}{2} x^{3/2}
+\frac{4\pi \alpha}{\eta} x^{5/2}} \; . 
\eeq

The presence of quantized topological configurations, like 
vortices, is a clear signature of the existence 
of an underlying superfluid order parameter, which plays the key role 
in all superfluid to normal phase transitions \cite{leggett,babaev}. 
Quantized  { vortices} have been found in superconductors \cite{cond}, 
superfluid helium \cite{helium} and in trapped degenerate 
gas \cite{vl,gas}. A vortex-lattice  structure 
has been created in a rapidly rotating dilute trapped 
{ Bose-Einstein condensate}. 
Here we study the vortex-lattice generation in a rotating disk-shaped 
trapped { bosonic superfluid} along the weak coupling 
to unitarity crossover, performing open-multi-processor-parallelized 
numerical simulations  \cite{CN,CN2} of the nonlinear Schr\"odinger equation of 
the superfluid order parameter \cite{vlfort}, { whose nonlinear 
term is proportional to the chemical potential of 
Eqs. (\ref{eq2}) and (\ref{co})}. 
{ Throughout this study, we will assume the
system to be an ideal superuid at zero Kelvin temperature}.
{ Not surprisingly,} 
comparing these results for vortex lattice with those 
obtained from the simple mean-field Gross-Pitaevskii  
equation \cite{gross}, i.e. by using 
\beq 
\mu(n,a)=4\pi n a \; , 
\label{pippo}
\eeq
we find that the latter may lead to a qualitatively 
incorrect description. See \cite{ueda,feder,aftalion} for previous 
numerical results with the { Gross-Pitaevskii} equation.
The properties of a rotating dilute { bosonic superfluid} and 
of the generated vortex lattice, and in particular 
the number of vortices and  root-mean-square  (rms) sizes, 
are very sensitive to the value of the scattering length 
through the bulk chemical potential $\mu(n,a)$. In other 
words, different forms of $ \mu(n,a)$ lead to widely 
different vortex-lattice structures. 
{ Our} numerical simulations can be quite useful 
for future experiments on the formation of quantized vortices 
in bosonic systems made of alkali-metal atoms. 

\section*{Results}

\subsection*{Superfluid hydrodynamics and nonlinear equations}
 
{\it General Consideration}: According to Tisza \cite{tisza} and 
Landau \cite{landau}, at zero temperature the equations 
of superfluid hydrodynamics for a dilute degenerate Bose gas are 
given by \cite{leggett,rmp}
\beqa
{\partial n\over \partial t} + 
{\boldsymbol \nabla} \cdot \left( n {\bf v} \right) &=& 0 \; , 
\label{hy1}
\\
{\partial {\bf v}\over \partial t} +  {\boldsymbol \nabla} 
\left[ {1\over 2}  v^2 + V({\bf r}) - 
{\nabla^2\sqrt{n} \over 2 \sqrt{n}} + \mu(n,a) \right] &=& { 0} \; , 
\label{hy2}
\eeqa
where $n({\bf r},t)$ is the superfluid density, 
${\bf v}({\bf r},t)$ the superfluid velocity, and 
$V({\bf r})$ is the external trapping potential. 
These equations of a { nonviscous,}  irrotational fluid 
describe accurately many collective properties 
of { a bosonic superfluid} \cite{leggett,babaev}. 
The quantum pressure term 
$-\nabla^2\sqrt{n}/(2\sqrt{n)}$, which is absent in the original 
equations of Tisza \cite{tisza} and Landau \cite{landau}, 
is indeed necessary to model surface effects \cite{rmp}. 
{ It is important to stress that, in general, 
at zero temperature a bosonic system is fully superfluid also 
if its condensate fraction can be small. This is the case 
of $^{4}$He liquid (where the effective range of the inter-atomic 
potential is of the order of the interparticle distance), 
whose zero-temperature condensate fraction is 
less than $10\%$ \cite{leggett}. In the case our dilute 
bosonic gas, characterized by an effective range of the interaction 
much smaller than the interparticle distance,  
the zero-temperature condensate fraction is close to $100\%$ 
in the weak-couling regime and it reduces to about $80\%$ at 
unitarity \cite{RS}.} 

As suggested by Onsager \cite{onsager}, Feynman \cite{feynman} 
and Abrikosov \cite{abrikosov}, superfluids have another 
amazing property: the circulation 
of the superfluid velocity field ${\bf v}({\bf r},t)$ around a 
generic closed path ${\cal C}$ is quantized, namely  \cite{fetter}
\beq 
\oint_{{\cal C}} {\bf v} \cdot d{\bf r} = 2\pi \ q \; , 
\label{circolo}
\eeq
where $q=0,\pm 1,\pm 2,....$ is an integer number. 
If $q \neq 0$ it means that inside the closed path ${\cal C}$ there are 
topological defects, and the domain where ${\bf v}$ is well 
defined is multiply connected. 
A simple example of topological defect is a quantized vortex line.
Nowadays quantized vortices are observed experimentally  
in type-II superconductors \cite{type}, in superfluid liquid 
helium \cite{helium2}, 
and in ultra-cold atomic gases \cite{leggett,babaev,fetter}. 
The quantization of circulation can be explained 
following the old intuition of London \cite{london} 
and assuming that the dynamics of superfluids is driven by 
a complex scalar field (for an in-depth discussion see \cite{babaev,fetter}) 
\beq 
\phi({\bf r},t) = |\phi({\bf r},t)| \ e^{i\theta({\bf r},t)} \; , 
\eeq
which satisfies the { nonlinear Schr\"odinger equation}  \cite{rmp}
\beq 
i {\partial \over \partial t} \phi ({\bf r},t)= \left[-{1\over 2}\nabla^2 
+ V({\bf r}) \right] \phi ({\bf r},t)+{ \mu(n,a)} \ \phi ({\bf r},t)
\label{eq31}
\eeq
with $n({\bf r},t) = N|\phi({\bf r},t)|^2, \int |\phi({\bf r},t)|^2 
d{\bf r}=1$ and the phase 
$\theta({\bf r},t)$ defines the superfluid velocity 
${\bf v}({\bf r},t) = {\boldsymbol \nabla} \theta({\bf r},t)$, 
where $N$ is the number of atoms.
 
In fact, under these assumptions, Eq. (\ref{eq31}) is equivalent 
to Eqs. (\ref{hy1}) and (\ref{hy2}) and the multi-valued angle variable 
$\theta({\bf r},t)$ gives rise to Eq. (\ref{circolo}). 
The complex field $\phi({\bf r},t)$ is the superfluid order 
parameter of our zero-temperature theory. { In other words, 
$\phi({\bf r},t)$ is the wavefunction of our single-orbital theory 
for a system that is fully superfluid.} Notice that 
for $\mu(n,a)=4\pi n a$, i.e. for a very weakly interacting 
{ bosonic superfluid}, the { nonlinear Schr\"odinger equation} 
reduces to the familiar { Gross-Pitaevskii} 
equation \cite{gross} characterized by a cubic nonlinearity. 
Remarkably, only very recently, 
vortex arrays in Fermi superfluids through the crossover 
{ from the weak-coupling situation with largely overlapping 
atomic pairs to the strong-coupling limit where composite 
bosons form and condense}, 
have been studied numerically by using a differential equation 
for the local order parameter, obtained by coarse graining 
the Bogoliubov-de Gennes equations \cite{strinati}.

{\it Bulk chemical potential}: 
The bulk chemical potential { of Eq. (\ref{eq2})}  
includes the effect of atomic 
interaction on the properties of the trapped dilute 
{ bosonic superfluid} 
governed by Eq. (\ref{eq31}). Different functional forms { of $f(x)$} 
may lead to widely different properties, specially, of the 
generated vortex lattice in a rapidly 
rotating trapped { bosonic superfluid}. 
Similar { crossover}  functions have been suggested and 
used mostly in the case of a Fermi gas \cite{gior,fermi}. 

\begin{figure}[!t]
\begin{center}
\includegraphics[width=0.9\linewidth,clip]{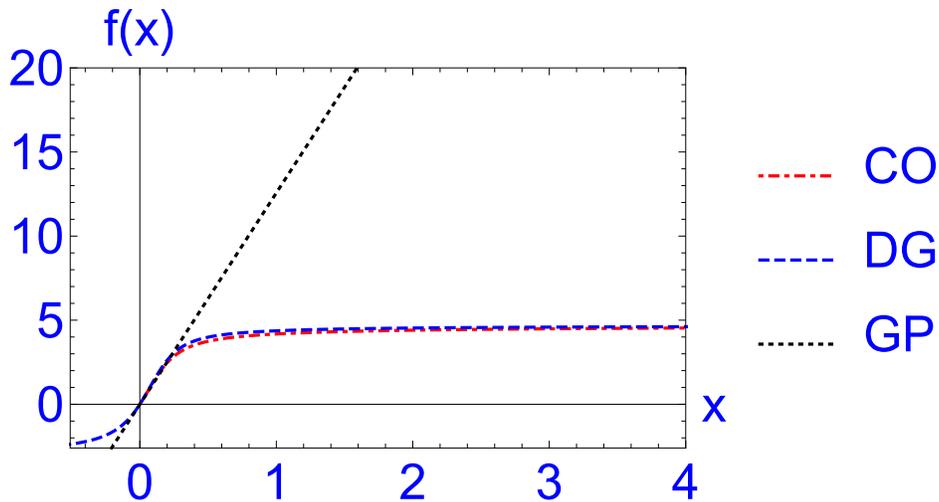}
\caption{Dimensionless function $f(x)$ of the 
zero-temperature bulk chemical potential $\mu=n^{2/3} f(x)$
versus $x=n^{1/3}a$. CO: the { crossover}  function (\ref{co}) 
with $\eta =4.7$,  DG: { multi-orbital} calculation of Ding and 
Greene \cite{DG}, GP: Gross-Pitaevskii function $f(x)=4\pi x$.}
\label{fig1} 
\end{center}
\end{figure} 

In Fig. \ref{fig1}  we display  the  { crossover} function (\ref{co}) for 
$\eta=4.7$, and compare it with that of the { Gross-Pitaevskii} 
model, { Eq. (\ref{pippo})},  
and the recent { microscopic multi-orbital Hartree} calculation of Ding 
and Greene \cite{DG}. We find  that the { crossover} function 
agrees well with the { multi-orbital Hartree} calculation 
for all $x$. The { Gross-Pitaevskii} function, 
$f(x)=4\pi x$ is reliable only for a relatively small value 
of the gas parameter $x$.  

{\it Rapidly rotating degenerate Bose gas}:  Abrikosov \cite{abrikosov} 
demonstrated from energetic consideration that a rapidly rotating 
Bose gas prefers many vortices of unit angular momentum per atom arranged 
in a regular lattice over a vortex of multiple angular momentum. 
It has been demonstrated \cite{ueda,feder} that   
the formation of such vortex lattice { in a trapped degenerate 
Bose gas} can be well described { by the following stationary 
dimensionless equation in the rotating frame obtained}
upon the  inclusion of  a term $-\Omega L_z$  in  Eq. (\ref{eq31})  \cite{llf}:
\begin{align}
\label{eq41}
i  \frac{\partial \phi({\bf r},t)}{\partial t} & =
\biggr[-\frac{1}{2} \nabla^2_{\bf r}  + 
\mu(n,a)+ i {\Omega} \left(x \frac{\partial }{\partial y}-
y \frac{\partial }{\partial x}               \right)\nonumber \\ 
&+ \frac{1}{2} (\gamma^2 x^2+\nu^2 y^2+\lambda^2 z^2)
\biggr]  \phi({\bf r},t) \; ,
\end{align}
{ where $\Omega $ is the frequency of rotation about $z$ axis }
  and $L_z$ is the  $z$ component of 
angular momentum: 
$L_z\equiv - i\hbar (x \frac{\partial }{\partial y}-y \frac{\partial }
{\partial x}          ). $
In  Eq. (\ref{eq41}) the   anisotropies $\gamma, \nu,\delta$ of the 
3D trap $V({\bf r})$ 
along $x,y,z$ axes, respectively,  are explicitly shown  in units of an 
overall frequency $\omega$ { of the confining trap}, the length 
is expressed in units of 
$l\equiv  \sqrt{\hbar/m\omega}, $  the angular frequency $\Omega$ is 
expressed in units of $\omega$, time $t$ in units of $\omega^{-1}$, 
$\mu$ in units of $\hbar \omega$, $\phi$ in units of $l^{-3/2}$. 
The stable vortex lattice emerges as the ground state of Eq. (\ref{eq41}), 
which we study numerically by imaginary-time propagation \cite{feder}. 

{\it Dimensional reduction}: Many experiments are performed in strongly 
disk- or cigar-shaped condensates.  The present { crossover} model 
leads to useful 
quasi two-dimensional (2D) or one-dimensional (1D) nonlinear models 
in these cases, which are included here for the sake of completeness.  
If we have a stronger trap in the $z$ direction ($\lambda \gg \nu, \gamma$), 
the quasi-2D nonlinear Schr\"odinger equation for the weak-coupling 
to unitarity crossover can be written as (viz. Methods for details):
 \begin{align}\label{q2d}
i \frac{\partial \phi_{2D}(x,y,t)}{\partial t } &= \Big[ -\frac{1}
{2}\Big(\frac{\partial^2}{\partial x^2}+ \frac{\partial^2}
{\partial y^2} \Big) + \mu_{2D}(n_{2D}, a) \nonumber \\
 & + 
 \frac{1}{2}(\gamma^2 x^2+\nu^2 y^2) 
 \Big]  \phi_{2D}(x,y,t),
\end{align}
with the normalization $\int dx dy | \phi_{2D}(x,y,t)|^2 =1$, where 
 \begin{align}
\mu_{2D}(n_{2D},a)=
\frac{4\pi  \frac{ {\sqrt{n_{2D}}}}{ {\sqrt{2 \pi}d_z} }  \left(x_{2D} + 
\frac{{2}\alpha  x_{2D}^2 \sqrt a}{\sqrt{5d_z} \pi^{1/4}}\right)  }
{1+ \frac{\alpha  x_{2D} \sqrt a}{\sqrt{5d_z} \pi^{1/4}}+\frac{{8} 
\alpha \pi^{7/12} }{\sqrt 6 \eta } \big( \frac{ax_{2D}^2}{d_z} \big)^{5/6}  },
\end{align}
with $d_z = l/\sqrt \lambda. $ Similarly, if we have a stronger trap in 
$x,y$ directions ($\gamma, \nu \gg \lambda$), the quasi-1D 
{ nonlinear Schr\"odinger} equation  
for the crossover is given by (details in Methods):
 \begin{align}\label{q1d}
i \frac{\partial \phi_{1D}(z,t)}{\partial t} = 
\Big[ -\frac{1}{2}\frac{\partial^2}{\partial z^2} + \mu_{1D}(n_{1D}, a) 
  + 
 \frac{\lambda^2 z^2}{2} 
 \Big]  \phi_{1D}(z,t), 
\end{align}
with the normalization $\int dz | \phi_{1D}(z,t)|^2 =1$, where    
 \begin{align}
\mu_{1D}(n_{1D},a)&=
 \frac{\frac{{2 }}{ {{d_x d_y }} }  \left( x_{1D}+ \frac{4\alpha  }
{5 \sqrt{\pi d_x d_y} }  a x_{1D}^{3/2}  \right) }{1
+ \frac{2\alpha  }{5 \sqrt{\pi d_x d_y} }  a x_{1D}^{1/2}
+ \frac{{ 8}\alpha \pi^{1/6}}{3 \eta (d_x d_y)^{5/6}}
 a^{5/3}  x_{1D}^{5/6} },   
\label{eq18}
\end{align}
with $d_x=l/{\sqrt \gamma}, d_y =l/{\sqrt \nu}$.
 
\subsection{Numerical Results}

The 3D, quasi-2D, and quasi-1D { crossover} equations 
(\ref{eq31}), (\ref{q2d}) 
and (\ref{q1d})   do  not  have    analytic  solution  and
different numerical methods, such as split-step Crank-Nicolson 
\cite{CN,CN2} and Fourier spectral
\cite{FS} methods, can be  used for their solution. 
In the following we undertake { the 
study of vortex-lattice formation  from a solution of} the 
relevant equations by the Crank-Nicolson method \cite{vlfort}.

\begin{figure}[!t]
\begin{center}
\includegraphics[width=0.55\linewidth,clip]{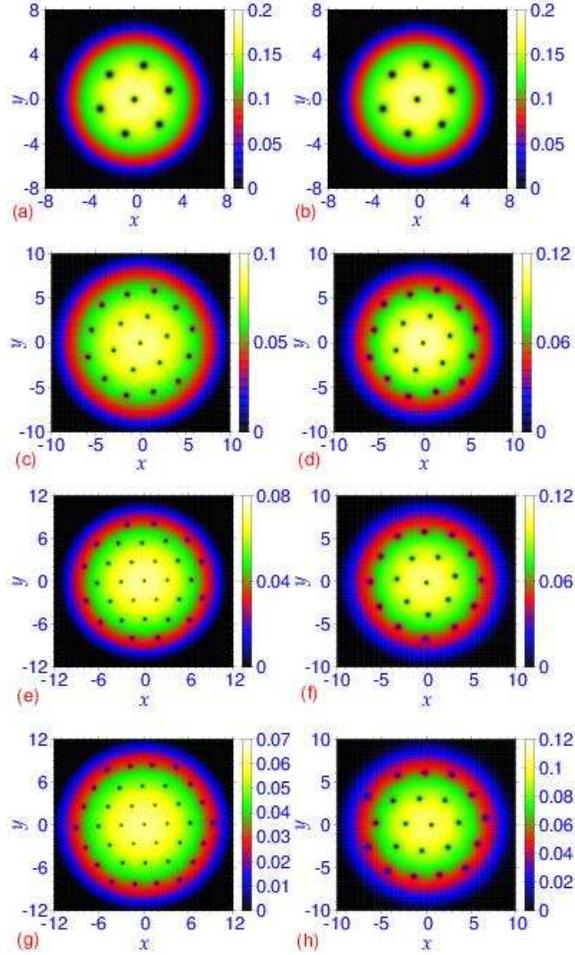}
\caption{The generated vortex lattice 
from a contour plot of the reduced 2D density in the $x$-$y$ plane 
$n_{2D}(x,y)$ from the { Gross-Pitaevskii (GP) model}, { i.e. 
Eq. (\ref{pippo})}, and the { crossover} (CO) model, 
{ i.e. Eq. (\ref{co})},  by using { the nonlinear 
Schr\"odinger equation} (\ref{eq41}) for (a) $ a=500a_0$ (GP), 
(b)  $ a=500a_0$  (CO),   
(c) $ a=2000a_0$  (GP), 
(d) $ a=2000a_0$  (CO),  
(e) $ a=3000a_0$  (GP),  
(f) $ a=3000a_0$  (CO), 
(g) $ a=4000a_0$ (GP),  
(h) $ a=4000a_0$ (CO).  
The other parameters of Eq. (\ref{eq41}) are 
$N=500, \Omega/\omega =0.4, \gamma=\nu=1, \lambda =900, \eta=4.7, 
l=1$ $\mu$m.  
The plotted quantities in this and following figures are dimensionless.  }
\label{vl}
\end{center}
\end{figure} 

In particular, we compare { the results from the } 
{ crossover model}, { given by 
the nonlinear Schr\"odinger equation (\ref{eq41}) 
with $\mu(n,a)$ obtained from Eq. (\ref{co})}, with 
the { Gross-Pitaevskii} equation, 
{ that is Eq. (\ref{eq41}) with Eq. (\ref{pippo})}. 
For this study  we take a trapped disk-shaped Bose gas of $500$ atoms in 
the $x$-$y$ plane  with    $\gamma =\nu =1, \lambda =900$. 
A large $\lambda $ makes a strongly disk-shaped Bose gas to generate 
stable vortex lattice  without transverse  instability \cite{aftalion} of 
the vortex lines.  A  small $\lambda$, on the other hand, leads to bent 
vortex lines \cite{aftalion} which may destroy a clean vortex-lattice 
structure.   Moreover, we take the oscillator length $l=1$ $\mu$m, which 
for a trapped Bose gas of $^{87}$Rb atoms corresponds to a trap frequency 
$\omega \approx 2 \pi \times 116$ Hz. The scattering length  will be taken 
as tunable to different { values  near a Feshbach 
resonance \cite{feshbach}.}  

From Fig. \ref{fig1} we see that the { plots of the universal 
function $f(x)$ versus $x$} of the 
{ Gross-Pitaevskii} and { the crossover models} 
separate at $x\approx 0.4$, where beyond 
mean-field corrections become important and 
this happens for scattering length $a >2000a_0$ in this case, where 
$a_0= 5.2917721067 \times 10^{11}$ m is the Bohr radius. 
In the strong coupling domain, the results for the observable of a trapped 
{ bosonic superfluid} as obtained from the { Gross-Pitaevskii} 
model and the { crossover} model will be different. 
In the case of most observables, such as density, rms
sizes, frequencies of oscillation \cite{osc} etc., this difference can be 
seen only after a careful comparison of the results.  However, the 
vortex-lattice structures of a rapidly rotating trapped 
{ bosonic superfluid} as obtained from the two models 
{ are} found to be qualitatively different 
with widely different number of vortices in the two cases. 

\begin{figure}[!t]
\begin{center}
\includegraphics[width=0.9\linewidth,clip]{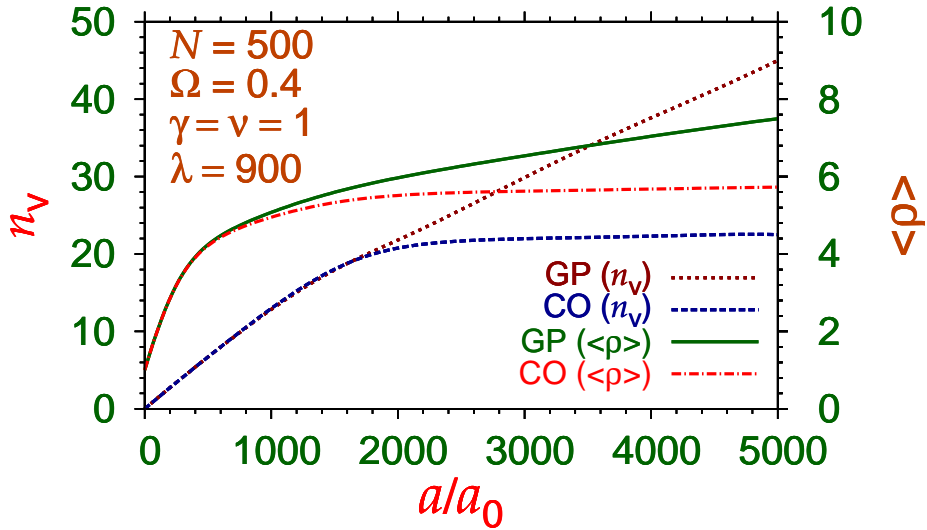}
\caption{Number of vortices $n_{\mathrm{v}}$ and the rms size 
$\langle \rho \rangle$ in the $x$-$y$ plane 
versus scattering length $a$ in a disk-shaped  trapped 
{ bosonic superfluid} of $N=500$ atoms rotating with angular 
frequency $\Omega=0.4$. 
Results obtained from the nonlinear Schr\"odinger equation (\ref{eq41}) 
using both the { Gross-Pitaevskii}  model,  
{ i.e Eq. (\ref{pippo})}, and { the crossover}  
model, i.e. Eq. (\ref{co}). The trap anisotropies are 
$\gamma=\nu=1, \lambda =900$ and the universal parameter  $\eta=4.7$.}
\label{fig3} 
\end{center}
\end{figure} 

To illustrate the vortex lattice, we plot in Fig. \ref{vl} the reduced 
2D density in the $x$-$y$ plane 
\begin{align}
n_{2D}(x,y) = \int dz |\phi(x,y,z)|^2
\end{align}
of a rapidly rotating trapped { bosonic superfluid} of 500 atoms 
with angular frequency 
$\Omega=0.4$ obtained from the solution of the { Gross-Pitaevskii} 
model and the { crossover} model 
for different scattering lengths $a$ ranging 
from $a=500a_0$ (weak coupling) to $a=4000a_0$ (strong coupling). 
In both the { Gross-Pitaevskii model} and the { crossover model}, 
for a fixed number of atoms $N$ and a fixed angular frequency $\Omega$, 
the number of vortices   increases as the scattering length increases 
resulting in a larger $ \mu(n,a)$: the linear system with $  \mu(n,a) =0$ 
does not generate vortex lattice independent of the rotational frequency. 
For $x>0.4$, the mean-field { Gross-Pitaevskii} model has larger 
nonlinearity than the beyond mean-field crossover model, viz. Fig. \ref{fig1}. 
Hence the { Gross-Pitaevskii} model generates 
more vortices than the { crossover} model in this domain 
for a fixed  { angular}
frequency $\Omega$, viz. Figs. \ref{vl}(a)-(h). From Fig. \ref{vl}  we find 
that the numbers of generated vortices for $a=500a_0, 2000a_0, 3000a_0,$ 
and $4000a_0$ using the { Gross-Pitaevskii model} 
({ crossover model}) 
are, respectively, 7 (7), 19 (19), 31 (20), 37 (22), demonstrating a 
saturation of this number in the { crossover} model.

In Fig. \ref{fig3} we plot the number of generated vortices 
$n_{\mathrm{v}}$ versus scattering length $a$ of the 
{ Gross-Pitaevskii model} and the { crossover} model. 
From Figs. \ref{vl} we find that the size of the 
trapped { bosonic superfluid} in the $x$-$y$ plane 
also increases with the scattering length. 
To quantify this difference between the two models, we also plot 
in this figure the rms sizes in the $x$-$y$ plane 
$\langle \rho \rangle$ of the two models. 
The number of generated vortices (and the rms size) in the 
{ Gross-Pitaevskii} model is always larger or equal to that 
in the { crossover} model. For smaller values of scattering length in 
the weak-coupling domain, 
the nonlinearities of both models are practically the same, and the number 
of generated vortices (and the rms size) in both models are the same. 
However, in the { Gross-Pitaevskii} model the number of vortices 
(and the rms size) increases indefinitely with the increase of 
scattering length, whereas the same in the { crossover} 
model { tends to saturate} for $a>2000 a_0$. 
 
\begin{figure}[!t]
\begin{center}
\includegraphics[width=0.7\linewidth,clip]{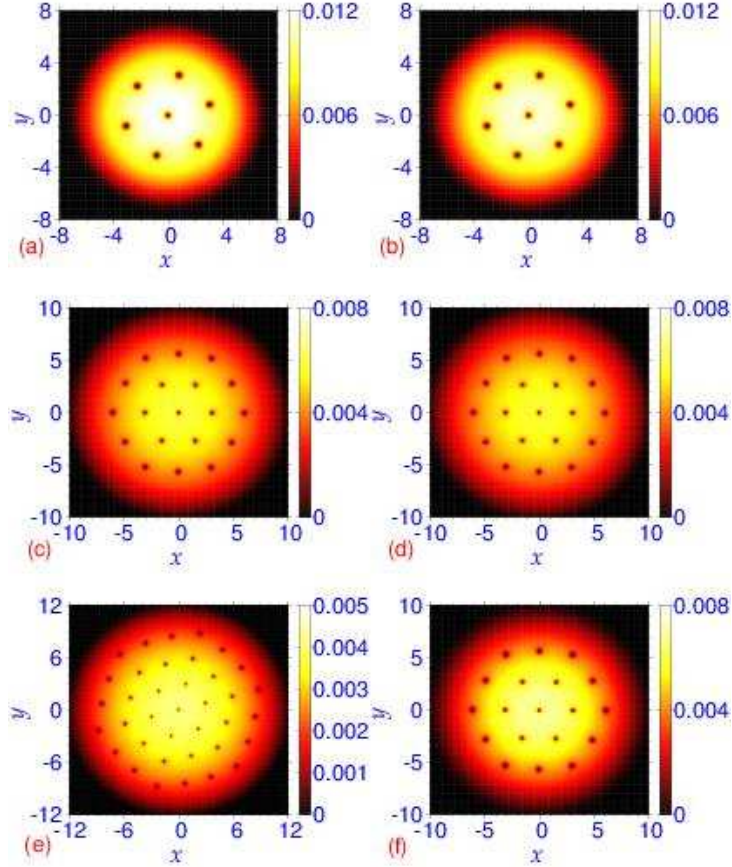}
\caption{The generated vortex lattice from a contour 
plot of the  2D density $|\phi_{2D}(x,y)|^2$ in the $x$-$y$ plane 
from the { Gross-Pitaevskii} (GP) model, 
{ given by Eq. (\ref{pippo})}  
and the { crossover} (CO) model, { given by Eq. (\ref{co})}, 
by using the { nonlinear Schr\"odinger equation} (\ref{q2d}) 
for (a) $ a=500a_0$ (GP), 
(b)  $ a=500a_0$  (CO),  
(c) $ a=2000a_0$  (GP), 
(d) $ a=2000a_0$  (CO), 
(e) $ a=4000a_0$  (GP), 
(f) $ a=4000a_0$  (CO). 
The other parameters are the same as in Fig. \ref{vl}.}
\label{fig4}
\end{center}
\end{figure} 

Next we use the quasi-2D model (\ref{q2d}) to see how well it can describe 
the vortex-lattice formation in the disk shaped trapped 
{ bosonic superfluid} studied above using the 3D model (\ref{eq41}) 
while employing the same sets of parameters in 2D and 3D. 
In Fig. \ref{fig4} we  { display}  the contour plots 
of the 2D densities for increasing scattering lengths: $a=500a_0, 2000a_0$, and 
$a=4000a_0$ using the quasi-2D Gross-Pitaevskii \cite{sala-gll} 
and { crossover} models (\ref{mu2d}). 
For the same value of scattering length,  the vortex lattice obtained in 
Fig. \ref{vl}  using a 3D description is quite similar to that obtained 
from the quasi-2D model in Fig. \ref{fig4}. For smaller values of scattering 
length ($a=500a_0, 2000a_0$), the quasi-2D  and the 3D models lead to  
identical description. For the larger scattering length $(a=4000a_0$), 
the number of vortices in the { crossover} model using  
3D and quasi-2D descriptions 
are 22 and 19, respectively. Considering the large value of 
the nonlinearity the agreement is quite fair: 
the quasi-2D model derived by integrating the transverse 
dynamics is valid for a small nonlinearity.

\section*{Discussion}

The commonly used { single-orbital} 
{ Gross-Pitaevskii} equation is appropriate to describe a 
trapped { pure Bose-Einstein condensate} for a small gas parameter $x$  
\cite{rmp}: small values of number of atoms and scattering length. 
Nevertheless, many experiments deal with a 
{ bosonic gases} with a larger gas parameter \cite{univ}, 
{ such that the zero-temperature condensate fraction is less 
than one \cite{leggett,RS} but the superfluid fraction is still 
one \cite{leggett,babaev}}. 
The { Lee-Huang-Yang} correction \cite{lee} to the 
standard { Gross-Pitaevskii} equation is appropriate to describe such a 
case. However, the { Lee-Huang-Yang} correction is not suitable for 
the strong-coupling unitarity limit for divergent values 
of the gas parameter. The properties of 
a { bosonic superfluid} in the unitarity limit show universal 
behavior \cite{univ} and hence is the topic of many investigations. 
We presented an analytic expression for the bulk chemical potential of 
a uniform Bose gas, suitable for the study of a trapped 
three-dimensional { bosonic superfluid} along the crossover
from weak-coupling to unitarity. This expression has the correct 
weak-coupling and unitarity limits in addition to the correct beyond 
mean-field { Lee-Huang-Yang} correction and faithfully reproduces
the results of an accurate microscopic { multi-orbital} 
calculation \cite{DG}. { Adopting} this analytic expression along the 
crossover, we study the evolution of vortex lattice 
in a rapidly rotating trapped { bosonic superfluid} 
along the weak coupling to unitarity 
crossover { by} using a 3D { single-orbital} 
{ nonlinear Schr\"odinger equation}, { which is equivalent 
to the generalized hydrodynamic equations of a superfluid 
at zero temperature}. { As expected, by increasing the 
interaction strength at { a} fixed { angular} frequency} 
we find dramatic differences { in} the vortex-lattice structures 
{ obtained}
from the { Gross-Pitaevskii} { model (\ref{pippo}) and the 
crossover model (\ref{co})}. Similar analytic expressions for the bulk 
chemical potential of quasi-1D and quasi-2D Bose gases are also derived
by integrating out the transverse dynamics.  
We also studied the vortex lattice in the quasi-2D model using the 
same parameters as the 3D model. For a strong transverse trap, 
the quasi-2D model provides a good description of the vortex lattice. 
{ We find that} the number  { of vortices from the 
Gross-Pitaevskii and the crossover models could}  be widely different: 
that in the { Gross-Pitaevskii} model 
diverges in the unitarity limit whereas in 
{ our crossover} model it remains finite consistent 
with the underlying physics. 
{ Thus, for future experiments with ultracold { Bose} 
gases made of alkali-metal atoms in the crossover 
from weak-coupling to unitarity, our numerical predictions 
based on the bulk equation of state (\ref{pippo}) and the 
superfluid nonlinear Schr\"odinger equation (\ref{eq31}) can be 
a quite useful reference.} 

\section*{Methods}

\subsection*{Quasi-1D configuration} 

If we have a stronger trap in the $x,y$ directions 
($\lambda \ll \gamma, \nu$), the dynamics in these  directions can be 
frozen to be confined in the harmonic-oscillator ground state of the 
trapping potential $m \omega^2(\gamma^2 x^2+ \nu^2 y^2)/2$:
\begin{align}
\Phi(x,y)= (\pi d_x^2)^{-1/4}   (\pi d_y^2)^{-1/4}e^{-x^2/2 d_x^2}  e^{-y^2/2 d_y^2},  
\end{align}
where $d_x =l/\sqrt \gamma,  d_y = l/ \sqrt \nu$, $l=\sqrt{\hbar/(m\omega)}$
and the wave-function is assumed to have the form \cite{sala-gll}
\begin{align}
\phi({\bf r},t) = \phi_{1D}(z,t)  \Phi(x,y),
\end{align}
where the effective dynamics is assumed to be confined only 
in the $z$ direction. 
The bulk chemical potential will now have the following 1D form \cite{sala-gll}
\begin{align}
\mu_{1D}(n_{1D},a)=\int  \mu(n,a) |\Phi(x,y)|^2 dx  dy,
\end{align}
where $n_{1D}= N |\phi_{1D}(z,t)|^2.$ Consequently, using Eq. (\ref{unit}) 
at unitarity, we obtain in the quasi-1D configuration
 \begin{align}\label{unit1D}
\lim_{a\to \infty} \mu_{1D}(n_{1D},a)=  \frac{ 3}
{ 5 (\pi \kappa) ^{2/3}}\eta n_{1D}^2,
\end{align}
where $\kappa = d_x d_y  n_{1D}^2$. The LHY correction (\ref{lhy}) in 
this case becomes
 \begin{align}\label{lhy1D}
\mu_{1D}(n_{1D},a)= 
\frac{2 n_{1D}^2}{ \kappa} \Big[ x_{1D}+\frac{2\alpha}{5} 
\frac{ x_{1D}^{5/2}}{\sqrt{\pi \kappa} }\Big],
\end{align}
where $x_{1D}= a {n_{1D}}$.  Equations (\ref{unit1D}) and (\ref{lhy1D}) 
lead to the  following quasi-1D chemical potential valid from weak coupling 
to unitarity:
 \begin{align}
\mu_{1D}(n_{1D},a)&= \frac{ {2 }}{ {{d_x d_y }} }
f_{1D}(x_{1D}), \\
f_{1D}(x_{1D})&=   \frac{ x_{1D} + \frac{4\alpha }{5}\frac{ x_{1D}^{5/2}}
{\sqrt{\pi \kappa} }} 
{1+  \frac{2\alpha}{5  } \frac{ x_{1D}^{3/2}}{\sqrt{\pi \kappa }}  +
\frac{{8}\alpha}{3 \kappa \eta} (\pi \kappa )^{1/6} x_{1D}^{5/2}} \label{eq17a} \\
  &= \frac{  x_{1D}+ \frac{4\alpha  }{5 \sqrt{\pi d_x d_y} }  a x_{1D}^{3/2}   }{1
+ \frac{2\alpha  }{5 \sqrt{\pi d_x d_y} }  a x_{1D}^{1/2}+ \frac{{ 8}
\alpha \pi^{1/6}}{3 \eta (d_x d_y)^{5/6}}
 a^{5/3}  x_{1D}^{5/6} },  
\label{eq18a}
\end{align}
where $f_{1D}(x_{1D})$ is dimensionless. 
All variables appearing in  Eq. (\ref{eq17a}) are dimensionless, 
but this expression is not appropriate for a numerical calculation 
as the denominator therein ($\kappa$) may have zero. 
In Eq. (\ref{eq18a}) this factor in the denominator 
has been cancelled to yield an expression  appropriate for 
a numerical calculation.

The quasi-1D nonlinear Schr\"odinger equation is then written as 
 \begin{align}
i \frac{\partial \phi_{1D}(z,t)}{\partial t} &= 
\Big[ -\frac{1}{2}\frac{\partial^2}{\partial z^2} + \mu_{1D}(n_{1D}, a) 
\nonumber \\
&  + 
 \frac{1}{2}m \omega^2\lambda^2 z^2 
 \Big]  \phi_{1D}(z,t),
\label{q1dnlse}
\end{align}
with the normalization $\int dz | \phi_{1D}(z,t)|^2 =1$.   
  
For a dilute { bosonic superfluid} with 3D scattering length $a$ 
and 1D number density $n_{1D} $, the {quasi-}1D weak-coupling regime 
{of cubic nonlinearity} 
is obtained if $a/(d_x d_y) \ll n_{1D} \ll 1/a$, while 
the  Tonks-Giarardeau regime {of quintic nonlinearity of a strictly 1D gas} 
holds if $n_{1D} \ll a/(d_x d_y)$ \cite{sala-gll}. 
It follows that the  { quasi-1D strong-coupling Bose gas cannot 
have the Tonks-Girardeau limit of a strictly 1D Bose gas.}  

\subsection*{Quasi-2D configuration}

If we have a stronger trap in the $z$ direction ($\lambda \gg \gamma, \nu$), 
the dynamics in this direction can be frozen to be confined in the 
harmonic-oscillator ground state of the trapping potential 
$m \omega^2\lambda^2 z^2/2$:
\begin{align}
\Phi(z)= (\pi d_z^2)^{-1/4} \exp(-z^2/2 d_z^2), \quad d_z =l/\sqrt \lambda,
\end{align}
where $l=\sqrt{ \hbar/m\omega}$.
and the wave-function is assumed to have the form \cite{sala-gll}
\begin{align}
\phi({\bf r},t) = \phi_{2D}(x,y,t)  \Phi(z),
\end{align}
where the dynamics is confined in the $x$-$y$ plane. 
The bulk chemical potential will now have the following 2D form
\begin{align}
\mu_{2D}(n_{2D},a)=\int  \mu(n,a) |\Phi(z)|^2 dz,
\end{align}
where $n_{2D}= N |\phi_{2D}(x,y,t)|^2.$ Consequently, using Eq. (\ref{unit}) 
at unitarity, we obtain in the quasi-2D configuration
 \begin{align}\label{unit2D}
\lim_{a\to \infty} \mu_{2D}(n_{2D},a)= \frac{\sqrt 3}
{\sqrt 5 \beta ^{1/3}}\eta n_{2D},
\end{align}
where $\beta =\pi d_z^2 n_{2D}$. 
The LHY correction (\ref{lhy}) in this case becomes
 \begin{align}\label{lhy2D}
\mu_{2D}(n_{2D},a)=4\pi  \frac{ {n_{2D}}}{ {\sqrt{2\beta }} } 
\Big[ x_{2D}+  \frac{\alpha x_{2D}^{5/2}}{\sqrt 5 \beta^{1/4}}\Big],
\end{align}
where $x_{2D}= a \sqrt{n_{2D}}$.  Equations (\ref{unit2D}) and (\ref{lhy2D}) 
yield following quasi-2D chemical potential valid from weak coupling 
to unitarity:
 \begin{align}\label{mu2d}
\mu_{2D}(n_{2D},a)&= \frac{ {\sqrt{n_{2D}}}}{ {\sqrt{2 \pi}d_z} }f_{2D}
(x_{2D}), \\
f_{2D}(x_{2D})&= 4 \pi \frac{ x_{2D} +  \frac{2\alpha x_{2D}^{5/2}}
{\sqrt 5  \beta^{1/4}}}
{1+  \frac{ \alpha x_{2D}^{3/2}}{\sqrt 5  \beta^{1/4}} + 
\frac{8 \pi \alpha x_{2D}^{5/2}}{\sqrt { 6} \beta ^{5/12} \eta } }, \\
&= 4\pi \frac{x_{2D} + \frac{{2}\alpha  x_{2D}^2 \sqrt a}{\sqrt{5d_z} 
\pi^{1/4}}  }{1+ \frac{\alpha  x_{2D} \sqrt a}{\sqrt{5d_z} \pi^{1/4}}
+\frac{{8} \alpha \pi^{7/12} }{\sqrt 6 \eta } 
\big( \frac{ax_{2D}^2}{d_z} \big)^{5/6} } .
\end{align}
The quasi-2D { Gross-Pitaevskii} model corresponds 
to $f_{2D}(x_{2D})=4\pi x_{2D}$. 
The quasi-2D nonlinear Schr\"odinger equation is then written 
as Eq. (\ref{q2d}), with the normalization 
$\int dx dy | \phi_{2D}(x,y,t)|^2 =1$.

\section*{Acknowledgements}

S.K.A. thanks the Funda\c c\~ao de Amparo \`a Pesquisa do Estado de 
S\~ao Paulo (Brazil) (Projects: 
2012/00451-0 and 2016/01343-7) and the Conselho Nacional 
de Desenvolvimento   Cient\'ifico e Tecnol\'ogico (Brazil) 
(Project: 303280/2014-0) for support. L.S. thanks for partial support 
the BIRD Project ``Superfluid properties of Fermi gases in 
optical potentials'' of the University of Padova { and 
the FFABR grant of Italian Ministry of Education, University and Research}. 

\section*{Author contributions}

The numerical part has been carried out by S.K.A. Analytical considerations
were performed S.K.A. and L.S. All the authors have contributed
to drafting the manuscript.

\section*{Additional information}

\textbf{Competing interests}: The authors declare no competing financial 
and non-financial interests.

\end{document}